\documentclass[11pt]{article}
\usepackage[utf8]{inputenc}
\usepackage{geometry}
\usepackage[natbib = true, sorting = none]{biblatex}
\usepackage{amsmath}
\usepackage{amssymb}
\usepackage{xcolor}
\usepackage{hyperref}
\usepackage{booktabs}
\usepackage{bm}
\usepackage{subcaption}
\usepackage{pdflscape}
\usepackage{setspace}
\hypersetup{colorlinks=true,linkcolor=blue,filecolor=blue,citecolor=blue}
\usepackage{tikz}
\usepackage{relsize}
\usepackage{multirow}

\title{A varying-coefficient model for characterizing duration-driven heterogeneity in flood-related health impacts\\[-1em]}
\author{Sarika Aggarwal, Phillip B. Nicol, Brent A. Coull, Rachel C. Nethery$^*$ \\
Department of Biostatistics, Harvard T.H. Chan School of Public Health\\
$^*$Corresponding author. Email: rnethery@hsph.harvard.edu
}
\date{}
\begin{document}

\maketitle

\begin{abstract}
\noindent Previous work revealed associations between flood exposure and adverse health outcomes during and in the aftermath of flood events. Floods are highly heterogeneous events, largely owing to vast differences in flood durations, i.e., flash-floods versus slow-moving floods. However, little to no work has incorporated exposure duration into the modeling of flood-related health impacts or has investigated duration-driven effect heterogeneity. To address this gap, we propose an exposure duration varying coefficient modeling (EDVCM) framework for estimating exposure day-specific health effects of consecutive-day environmental exposures that vary in duration. We develop the EDVCM within an area-level self-matched study design to eliminate time-invariant confounding followed by conditional Poisson regression modeling for exposure effect estimation and adjustment of time-varying confounders. Using a Bayesian framework, we introduce duration- and exposure day-specific exposure coefficients within the conditional Poisson model and assign them a two-dimensional Gaussian process prior to allow for sharing of information across both duration and exposure day. This approach enables highly-resolved insights into duration-driven effect heterogeneity while ensuring model stability through information sharing. Through simulations, we demonstrate that the EDVCM out-performs conventional approaches in terms of both effect estimation and uncertainty quantification. We apply the EDVCM to nationwide, multi-decade Medicare claims data linked with high-resolution flood exposure measures to investigate duration-driven heterogeneity in flood effects on musculoskeletal system disease hospitalizations.
\end{abstract}
\vspace{0.5em}

\section{Introduction}
In recent decades, flooding has affected more people worldwide than any other climate-related hazard \citep{hu2018flood}. Due to projected continuing increases in the intensity and frequency of floods, it is critical to evaluate the broad spectrum of health risks they impose \citep{esd_trends_2018,alfieri_warmer_2017,ceola_floodplains_2014}. Recent studies have established increased mortality and increased rates of hospitalization for several causes including musculoskeletal system diseases, nervous system diseases, skin diseases, and injuries during and in the immediate aftermath of flood exposure \citep{yang2023mortality,lynch2025large,aggarwal2025severe}. 

Floods are highly heterogeneous events and this heterogeneity is largely driven by the duration of the flood. Some floods are sudden and fast-moving (so-called ``flash floods''), with durations of a few hours or less. In other cases, slow-moving or stalled weather systems can bring excessive rain over long periods, leading to floodwaters that do not recede for several days to several weeks. No research to date has investigated how these different types of floods, characterized by their duration, may impact health differently. This is likely attributable to the dearth of methods that allow for investigating duration-driven heterogeneity in exposure effects within the time series models commonly used in climate epidemiology applications. Beyond floods, similar challenges arise when studying the health effects of other climate-related events with variable duration, such as heat waves and wildfires, underscoring the broader need for methods that can accommodate duration-dependent effects within standard modeling frameworks.

Motivated by this gap, this paper formulates a varying-coefficient modeling framework for estimating exposure day-specific health effects of consecutive-day environmental exposures that vary in length. In doing so, the proposed framework enables rigorous investigation of two scientific questions of interest in flood epidemiology. First, do effects of flood events vary with the length of the flood event? For example, do longer floods have more adverse effects than shorter floods? Second, does the length of the flood event result in different critical windows of vulnerability? For example, are adverse effects more concentrated during the later days of longer floods as compared to the earlier days of shorter floods? We also consider these questions in the context of lagged effects that follow immediately after the flooded days. These findings will have important implications for understanding the mechanisms underlying flood-related public health risks and response planning as well as the development of targeted flood adaptation strategies.

Building upon a self-matched study design and conditional Poisson regression approach commonly used to estimate health effects of climate events, we propose a model with duration- and day-specific exposure coefficients collectively assigned a two-dimensional Gaussian process prior that allows for sharing of information across durations and exposure-days. 
Specifically, after self-matching flood-exposed communities to themselves on the same days in prior and subsequent years (which act as controls), we model the expected daily rate of health outcomes in a community
as a function of binary flood exposure with a coefficient that varies based on both the duration of the flood and the current number of consecutive days of exposure. Due to the likelihood of sparse data for some exposure durations, information-sharing via the Gaussian process prior on the coefficients is crucial for model stability. This approach yields highly-resolved insights while ensuring model reliability and avoiding the need to arbitrarily discretize flood durations. We refer to our proposed approach as an exposure duration varying-coefficient model (EDVCM). 
The EDVCM is tested in simulations and applied to a dynamic Medicare cohort in the contiguous United States during 2000-2016, linked with high-resolution satellite-derived flood exposure data. We examine duration-driven heterogeneity in the associations between hospitalization and flood exposure for musculoskeletal system diseases which have previously been linked to flooding \citep{aggarwal2025severe}. 

\subsection{Prior work}\label{sec:priorwork}

In case-crossover study designs, which are common in environmental epidemiology, individuals who experience the health event of interest have their ``event day'' matched to nearby days without an event, and the relationship between short-term exposures and the health outcome are assessed in the matched data. Case-crossover studies, which are traditionally analyzed using conditional logistic regression treating an event day and its matches as a stratum, are a frequently-used alternative to standard time series regression methods (although in some special cases the two are equivalent). They are advantageous because individuals act as their own controls, eliminating time-invariant confounders. Choosing bi-directional controls (before and after the event) can help adjust for seasonal and secular trends \cite{jaakkola2003case}. 

Self-matched designs can also be executed at the area-level where instead of matching \textit{individuals} to themselves on other days, one matches a spatial unit, such as a ZIP Code or county, to itself on other days. Here, matching days are selected based on exposure rather than outcome, i.e., areas exposed to the climate event of interest have their exposure days matched to other days-- usually the same days-of-year in other years to account for seasonal trends. Area-level self-matched studies, which typically consider area-level rates of a health event as the outcome, are often analyzed with conditional Poisson regression models. 
Bayesian conditional Poisson regression models have been developed, including those with independent Gaussian priors on exposure coefficients \citep{barrera2023conditional}. However, to our knowledge, past approaches have only allowed estimation of effects for exposures of a common duration. Some prior work investigating exposures of differing durations ``collapsed'' exposure days together to accommodate them within this modeling framework \citep{aggarwal2025severe}.

Varying-coefficient models (VCM) extend classic linear models by allowing regression coefficients to vary smoothly as a function of covariates or ``effect modifiers'' \citep{hastie1993varying}. Each coefficient is modeled as a function of an effect modifier, such as time or location, allowing the association between exposure and outcome to change in magnitude or direction as the value of the effect modifier changes. This is particularly useful in settings where exposure-outcome relationships are not expected to be constant, including in longitudinal and spatio-temporal studies. Methodological developments have expanded their use through spline bases, kernel smoothing, and Bayesian formulations \citep{fan1999statistical, wu2000kernel, biller2001bayesian}. A particular VCM that has been widely used in environmental epidemiology is the distributed lag model (DLM). The typical DLM considers how the exposure at preceding time points, called lags, is associated with the response at a given time \citep{pope1996time, schwartz2000distributed, zanobetti2000generalized, braga2002effect}. This allows the model to capture potentially delayed (or ``lagged'') effects of environmental exposures. DLMs are most commonly used to study how the health impacts of exposures with substantial day-to-day variability, such as air pollution and temperature, unfold over time. 

Bayesian extensions of DLMs enable information sharing across locations and constrain the distributed lag function using prior knowledge \citep{peng2009bayesian, welty2009bayesian}, including the use of Gaussian process priors. Additional work has been done to characterize effect heterogeneity while simultaneously identifying windows of vulnerability using functional linear regression \cite{wilson2017bayesian} or Bayesian additive regression trees \cite{mork2023heterogeneous, mork2022treed}. Recent work has introduced distributed lag interaction models that allow the lag–response function to be modified by an index formed from multiple sociodemographic or environmental modifiers, with estimation proceeding in a Bayesian framework \citep{demateis2025distributed}. Chang et al. formulated a Bayesian hierarchical model that allows for distributed exposures to vary in two dimensions so that the exposure effect and the risk of the outcome can change over the course of the study \cite{chang2015assessment}.

While DLMs quantify the delayed health impacts of an exposure on a particular day, they can only provide limited insight into the impacts of exposure to discrete climate events that span multiple, consecutive days. For exposures of varying duration, such as floods, health impacts may depend both on lagged effects and on the number of consecutive days exposed. Our approach is designed specifically to account for this inherent variability in flood exposure duration by parameterizing the effect of each day of the event, while still allowing for post-flood (lagged) effects to be incorporated within the same modeling framework. Building on the VCM framework, we propose an approach that allows for coefficients to vary jointly across two dimensions (duration and exposure day), with one dimension (exposure day) constrained by the other (duration), and we allow for information-sharing using a Gaussian process prior. 

The remainder of this paper is organized as follows. Section 2 describes the motivating dataset which combines information from the Global Flood Database and Medicare claims data. Section 3 presents the EDVCM. In Section 4, we discuss a simulation study that evaluates the EDVCM's performance. Results from the real data analysis are given in Section 5. Finally, a discussion and future work are described in Section 6.

\section{Motivating data}\label{sec:data}
Hospitalization records were obtained from Medicare claims data from the Centers for Medicare and Medicaid Service \citep{cms}. We collected data on Medicare beneficiaries, aged 65 years or older, from January 1, 2000 to December 31, 2016 and living in the contiguous United States. From each billing claim, we extracted the beneficiary's county of residence, date of admission, and primary diagnosis ICD-9-CM code (ICD-10-CM code on or after October 1, 2015). ICD-9-CM and ICD-10-CM codes were grouped using the Clinical Classifications Software algorithm into broad causes of hospitalization \citep{elixhauser2014clinical}. Our outcomes of interest are (separate) county-level rates of hospitalization due to musculoskeletal system diseases.
\begin{figure}[h!]
    \includegraphics[width=1\textwidth]{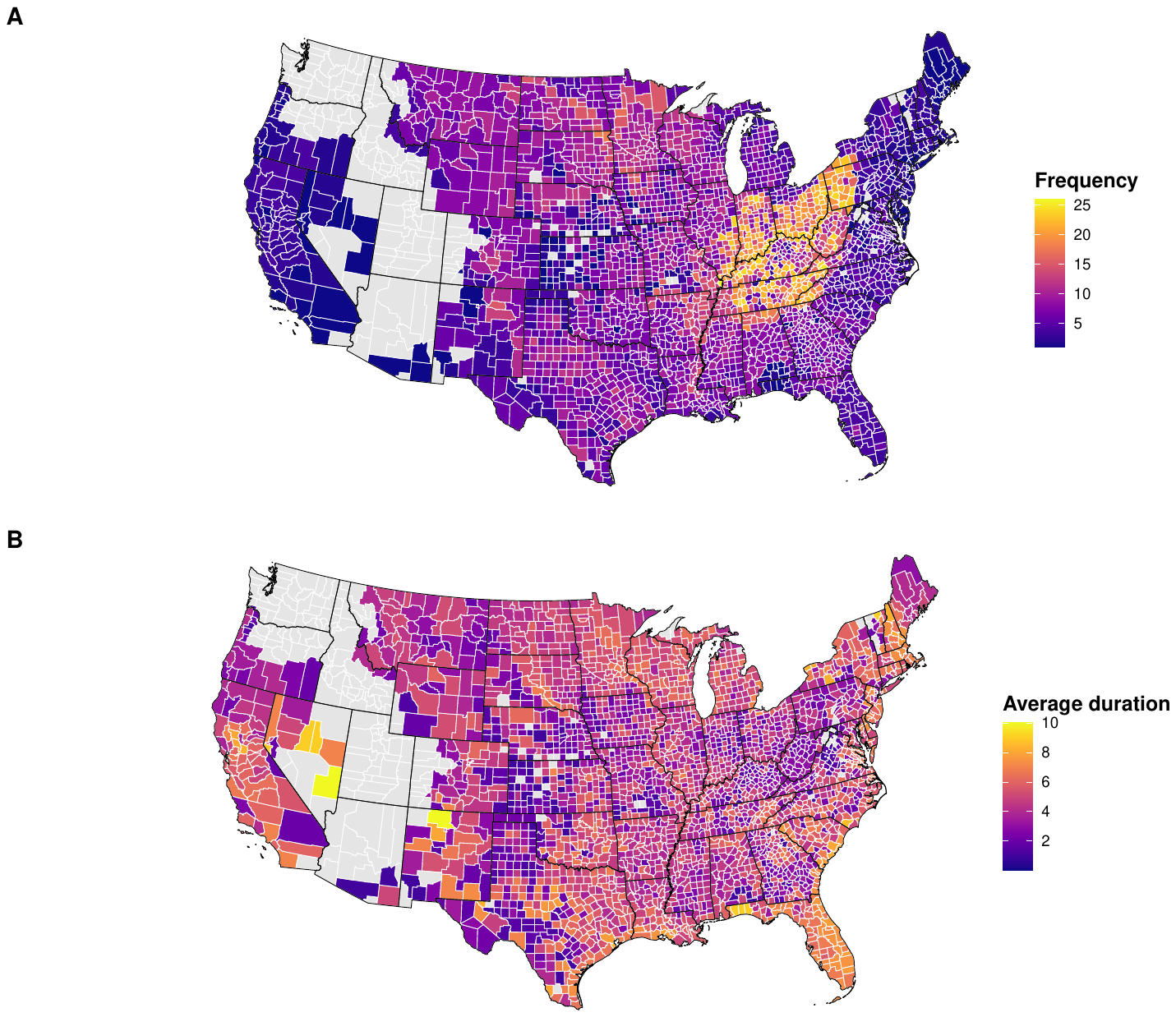}
    \caption{County-level flood characteristics in the contiguous United States from 2000–2016 for the data application. Panel A shows the number of flood events by county and Panel B shows the average flood duration (in days) by county. State borders are shown in black while county borders are shown in white. }
    \label{spatial_trends}
\end{figure}

We obtained data on severe flood exposures that occurred in the contiguous United States from 2000-2016 using the Global Flood Database, the most comprehensive collection of satellite-based high-resolution historic flood maps to date \citep{tellman_satellite_2021}. These flood events covered 47 states and the District of Columbia (Figure \ref{spatial_trends}, Panel A). We used the high-resolution maps to identify counties and days where floods occurred in the contiguous United States. To link flood exposure information with our health records, we aggregated it to the county level \citep{flood_data}. We defined a county-day as flood-exposed if the flood maps indicated any flooding (i.e., at least one flooded pixel within the county on that day). For counties classified as flood-exposed, we used the flood maps to obtain the number of days each pixel within the county was flooded during the event, and we defined the county's flood duration as the maximum of these pixel values. Given each exposed counties' location, start date, and duration, we created a row in the analytic data for every county-day of the flood event (for all events), which we then merged with the county-level Medicare hospitalization data.

We utilize an area-level self-matched study design as described in Section~\ref{sec:priorwork}. Specifically, we matched each flood-exposed county-day to a non-flooded day in the same county and on the same day-of-year but in years preceding or following flood exposure. Matched days were selected from the closest two years with no flood exposure during the same days-of-year or in the four weeks that followed. We excluded the four weeks post-flood from the control pool to allow for potentially delayed effects of flood exposure. In particular, the EDVCM accounts for these lagged (post-flood) effects by including lag indicators that capture health impacts occurring in the days following each flood event. The analytic dataset consists of all flood-exposed county-days, their matched controls, and the corresponding lag days following each of the exposed and control periods. 

While our self-matched study design accounts for time-invariant confounders, we must formally adjust for relevant time-varying covariates, including meteorological features and air pollution concentrations. We obtained daily 4-km gridded maximum air temperature, maximum relative humidity, and wind velocity data from the Gridded Surface Meteorological (gridMET) Dataset \citep{gridmet}. We also utilized daily 1-km gridded predictions of ambient fine particulate matter (PM2.5), ozone (O3), and nitrogen dioxide (NO2) pollutant concentrations \citep{pm25_pub, no2_pub, ozone_pub}. Gridded measures were aggregated to the county level using an unweighted average of all grid-cell values within each county \citep{gridmet_harvard, particulate_matter, nitrogen_dioxide, ozone}. 

\section{Methods}

\subsection{Exposure duration varying-coefficient model}

Units (county–day observations for a particular flood event) in the analytic data are indexed by $i$. Each unit $i$ is nested within a single stratum where strata are defined as the set of all exposed days, lag days, and matched controls for a given county flood exposure event. We let $s(i)$ be a function that returns the stratum index for unit $i$. Each stratum also corresponds to a flood exposure of a particular duration, and that flood duration is returned by a function $d(i)$. Then, for each unit representing a flood exposure time point (as opposed to a lag day), the corresponding time point (here, day) is returned by the function $t(i)$. Note that $d(i)$ and $t(i)$ are integer-valued and $t(i) \leq d(i)$ since exposure days cannot extend beyond the length of the event. Therefore, for exposure days, $t(i) \in \{1,2,\dots, d(i)\}$. Similarly, $l(i) \in \{1,2,\dots\}$ represents lag days, which are also integer-valued. By construction, the indexing functions $t(i)$ for exposure days and $l(i)$ for lag days are mutually exclusive, i.e., $t(i)$ is only defined if and only if $l(i)$ is not. For matched control units, $t(i)$ returns the exposure-day index corresponding to the matched control day and $l(i)$ returns the lag-day index corresponding to the matched control lag day, depending on which is defined. We let $D$ denote the maximum duration of all flood events in the analytical sample. 

Let $Y_{i}$ denote the hospitalization count for unit $i$ and $P_i$ denote the person-time denominator. Let $A_i$ denote the binary flood exposure status of unit $i$. For unit $i$, $L_{i}$ is a vector of indicators of which, if any, post-flood lag day unit $i$ corresponds to and $Z_i$ is the vector of measured time-varying covariates. 

We formulate our EDVCM within a Poisson regression modeling framework, although it could be integrated into many standard environmental epidemiology modeling approaches. Specifically, we assume $Y_i\sim \text{Poisson}(\lambda_i)$ where
\begin{equation}\label{poisson_model}
    \begin{split}
        \log \left(\lambda_i \right)  &= \alpha_0 + \alpha_{s(i)} + \beta_{d(i),t(i)} A_{i} + \theta'_{d(i),l(i)}L_{i} + \zeta'Z_{i} + \log(P_i)
    \end{split}
\end{equation} 
Here,
the $\alpha_{s(i)}$ are stratum-specific intercepts, which will later be conditioned out.  The $\beta_{d(i),t(i)}$ are the duration- and day-specific flood-attributable changes in the expected hospitalization rate on the log scale. These parameters are our primary quantities of interest. The $\theta_{d(i),l(i)}$ are the flood-attributable changes in the expected rate of hospitalization (on the log scale) on lagged days following exposure for a flood of duration $d(i)$. $\zeta$ is a vector of regression coefficients corresponding to the vector of time-varying covariates. In practice, the coefficient estimates from this model are typically exponentiated to allow interpretation as rate ratios.

In matched study designs, a model that includes stratum-specific intercepts, like our $\alpha_{s(i)}$ in Model \ref{poisson_model}, is often referred to as a ``conditional'' model. These are typically considered to be nuisance parameters. In frequentist versions of conditional logistic and conditional Poisson regression, equivalent results (for other model parameters) are obtained in models that estimate the stratum-specific intercepts and alternative formulations that eliminate them from the likelihood by conditioning on sufficient statistics (the sum of the outcomes in the stratum). While the two constructions give equivalent results in the frequentist setting \citep{lancaster2002orthogonal}, this may not be the case in the Bayesian setting. This is because, when the stratum-specific intercepts are kept in the model and assigned prior distributions, subtle shrinkage and information-borrowing issues might result in deviations from results of a model that, instead, conditions out these parameters. Moreover, since the number of stratum-specific coefficients will often be large, eliminating them can substantially reduce computational burden. Therefore, we reformulate Model~\ref{poisson_model} to condition out the stratum-specific intercepts, yielding a multinomial likelihood.

We let $\Psi(s)$ denote the set of all units in stratum $s$. Correspondingly, $\boldsymbol{Y}_{\Psi(s)}$ is the vector of hospitalization counts for all units in stratum $s$. Then, it is a well-known property of independent Poisson random variables that 
\begin{equation}\label{multinomial_fact}
\left(\boldsymbol{Y}_{\Psi(s)} | \sum_{i \in \Psi(s)} Y_{i} = W_s\right) \sim \text{Multinomial}\left(W_s, \boldsymbol{\pi}_{\psi(s)}\right)
\end{equation}
where $W_s$ is the sum of the hospitalizations across all units in stratum $s$ and $\boldsymbol{\pi}_{\psi(s)}$ is a vector of probabilities of assigning each of the $W_s$ hospitalizations to each of the units in the stratum. The probability parameter for each unit has the following connection to the parameters from Model~\ref{poisson_model}:
\begin{align*}\label{multinomial_probability}
\pi_{i} \ = \ \frac{\lambda_{i}}{\sum_{i' \in \Psi(s)} \lambda_{i'}} \ = \ \frac{\text{exp}({\beta_{d(i),t(i)} A_i + \theta_{d(i),l(i)} L_i + \zeta'Z_i}) \times P_i}{\sum_{i' \in \Psi(s)}\text{exp}({\beta_{d(i'),t(i')} A_{i'} + \theta_{d(i'),l(i')} L_{i'} + \zeta'Z_{i'}}) \times P_{i'}}
\end{align*} and 
\begin{equation*}\label{multinomial_model}
        \log \left(\pi_{i} \right)  = \beta_{d(i),t(i)} A_i + \theta_{d(i),l(i)}L_i + \zeta'Z_i + \log(P_i) 
        - \log \left(\sum_{i' \in \Psi(s)} \text{exp}({\beta_{d(i'),t(i')} A_{i'} + \theta_{d(i'),l(i')}L_{i'}+\zeta'Z_{i'}})\times P_{i'} \right)
\end{equation*} where the final term serves as a non-traditional offset that constrains the multinomial probabilities to sum to one.

\subsection{Multi-dimensional coefficient smoothing via Gaussian process priors}
Model \ref{poisson_model} allows for flood effects to vary across length of the flood event and days within the flood event through the coefficients $\beta_{d(i),t(i)}$. This applies analogously to the lagged flood effects, $\theta_{d(i),l(i)}$. In this section, we denote the coefficients for day $t$ of a flood of duration $d$ by $\beta_{d,t}$, dropping the functional subscripts which are used in the full model specification only to map units to their corresponding day and duration (similarly for $\theta_{d,l}$). In total, there are $1 + 2 + \ldots + D = \frac{D(D+1)}{2}$ day- and duration-specific flood exposure coefficients which can be visualized in the lower triangular matrix below
\begin{equation}\label{beta_matrix}
\boldsymbol{\beta}=\begin{bmatrix}
\beta_{1,1} \\
\beta_{2,1} & \beta_{2,2} \\
\beta_{3,1} & \beta_{3,2} & \beta_{3,3} \\
\vdots & \vdots & \vdots & \vdots & \vdots & \dots & \ddots \\
\beta_{D-1,1} & \beta_{D-1,2} & \dots & \dots & \dots & \dots & \beta_{D-1,D-2} & \beta_{D-1,D-1} \\
\beta_{D,1} & \beta_{D,2} & \dots & \dots & \dots & \dots & \beta_{D,D-2} & \beta_{D,D-1} & \beta_{D,D} \\
\end{bmatrix}
\end{equation}, where each row contains all coefficients for floods of a given duration and columns contain the coefficients for a particular exposure day across all durations. 

In order to estimate the $\beta_{d,t}$, the EDVCM borrows information across flood durations and days within the flood event by assigning a Gaussian process (GP) prior structure. While our analytic dataset is large, daily county-level hospitalization counts can be small and there are few floods that persist for longer durations, so we have limited data to inform some of the $\beta_{d,t}$ parameters. Consequently, sharing information across exposure days and durations is necessary to stabilize estimation when estimating $\boldsymbol{\beta}$. In addition to stabilizing coefficient estimation, this is also motivated by prior knowledge that the health impacts of flood exposures are likely to vary smoothly across flood durations and exposure days. It is also possible that we may have no observed data corresponding to particular duration(s) within $[1,D]$. In such cases, the smoothing induced by the GP prior allows for the corresponding row of $\boldsymbol{\beta}$ to be (potentially weakly) identified. 

We let $\tilde{\boldsymbol{\beta}}$ denote the $\frac{D(D+1)}{2}$-length vector of the coefficients contained in the matrix $\boldsymbol{\beta}$. The GP prior for $\tilde{\boldsymbol{\beta}}$ is in the form of $\tilde{\boldsymbol{\beta}} \sim MVN\left(\boldsymbol{0}, \boldsymbol{\Sigma}_\beta \right)$. $\boldsymbol{\Sigma}_\beta$ is a $\frac{D(D+1)}{2}\times \frac{D(D+1)}{2}$ covariance matrix with the entry corresponding to the covariance between $\beta_{d,t}$ and $\beta_{d',t'}$ given by
\begin{equation}\label{beta_kernel}
    \begin{split}
    Cov(\beta_{d,t}, \beta_{d',t'}) &=  \sigma^2_{\beta} \text{exp}\left(-\frac{1}{\phi}|d - d'| - \frac{1}{\tau}|t-t'|\right) \\
    &= \sigma^2_{\beta} \text{exp}\left(-\frac{1}{\phi}|d - d'|\right)\text{exp}\left(-\frac{1}{\tau}|t-t'|\right)
    \end{split}
\end{equation} 
, which is the product of exponential kernels for each dimension (duration and exposure day). This multi-dimensional product kernel \cite{duvenaud2014automatic} allows for two-dimensional smoothing across days of flood exposure ($t$) and the duration of flood event ($d$). Each input dimension, $d$ and $t$, has a different lengthscale parameter, $\phi$ and $\tau$, respectively and we assume a single variance parameter, $\sigma^2_{\beta}$. The lengthscale parameters, which are assigned hyperpriors and learned from the data, determine the degree of smoothness/information sharing across each dimension. By specifying a separable product kernel, we induce smoothness across exposure days and durations independently, prioritizing interpretability and stability over more complex interaction surfaces. Under this kernel structure, more information is shared in the estimation of $\beta_{d,t}$ and $\beta_{d',t'}$ if $d$ is close to $d'$ (floods of similar length) and $t$ is close to $t'$ (consecutive or nearby days within a flood event). This prior specification is conceptually related to generalized additive models (GAMs), where smoothness is imposed marginally along each input dimension, while retaining the flexibility of a Bayesian hierarchical framework for uncertainty quantification. 

Let $\boldsymbol{\theta}$ be the $D\times l$ matrix of lagged coefficients where $\tilde{\boldsymbol{\theta}}$ is the corresponding vector representation. We utilize an analogous prior structure for $\tilde{\boldsymbol{\theta}}$ such that $\tilde{\boldsymbol{\theta}} \sim MVN\left(\boldsymbol{0}, \boldsymbol{\Sigma}_\theta \right)$ where $\boldsymbol{\Sigma}_\theta$ has the form of 
\begin{equation}\label{theta_kernel}
\begin{split}
    Cov(\theta_{d,l}, \theta_{d',l'}) &=  \sigma^2_{\theta} \text{exp}\left(-\frac{1}{\gamma}|d - d'| - \frac{1}{\eta}|l-l'|\right) \\
    &= \sigma^2_{\theta} \text{exp}\left(-\frac{1}{\gamma}|d - d'|\right)\text{exp}\left(-\frac{1}{\eta}|l-l'|\right)
\end{split}
\end{equation}.
This formulation of the GP prior for $\boldsymbol{\theta}$, similarly, allows for smoothing of lagged effects to depend both on the closeness of lag days as well as duration of the flood event. However, if lagged effects are not believed to be a function of the length of the flood event, then it is straightforward to smooth only in a single dimension. Additionally, we allow for the lengthscale parameter for duration in the kernel for the lagged parameters ($\gamma$) to differ from that of the primary parameters of interest ($\phi$), but note that $\gamma$ in \ref{theta_kernel} can be replaced by $\phi$. 

\subsection{Estimation and summary measures}

We use weakly informative priors on all hyperparameters ($\sigma^2_{\beta}, \phi, \tau, \sigma^2_{\theta}, \gamma, \eta$) to allow the data to determine the degree to which scaling and smoothing occurs (Appendix \ref{hyperpriors}). We then use Hamiltonian Monte Carlo sampling to collect samples from the posterior distribution of all parameters. We report posterior means as point estimates and construct 95\% credible intervals using the percentile method.

While a strength of the EDVCM approach is that it allows for high-resolution characterization of flood-related health impacts, a cumulative effect measure may be useful in some contexts to summarize the total health impact of floods for a given duration. It can be shown that when there is no model-based adjustment of time-varying covariates, the cumulative rate ratio for floods of a particular duration is the average of the exponentiated betas: 
$$\Delta_{d} = \frac{1}{d}\sum_{\substack{i:\,d(i)=d, \\ t(i) \leq d}}\text{exp}(\beta_{d(i),t(i)})$$
With time-varying covariates, the cumulative rate ratio is $$\Delta_{d} = \frac{\sum_{i:d(i)=d, t(i) \leq d}\text{exp}(\beta_{d(i),t(i)})\text{exp}(\zeta'z_{i})}{\sum_{i:d(i)=d, t(i) \leq d}\text{exp}(\zeta'z_{i})}$$ Mathematical details are provided in the Appendix (\ref{cumulative_deriv}). Posterior samples of the cumulative rate ratios are constructed from the posterior samples of the individual parameters.

\section{Simulation study}\label{simulation_sec}
\begin{figure}[h!]
    \centering
    \includegraphics[width=1\textwidth]{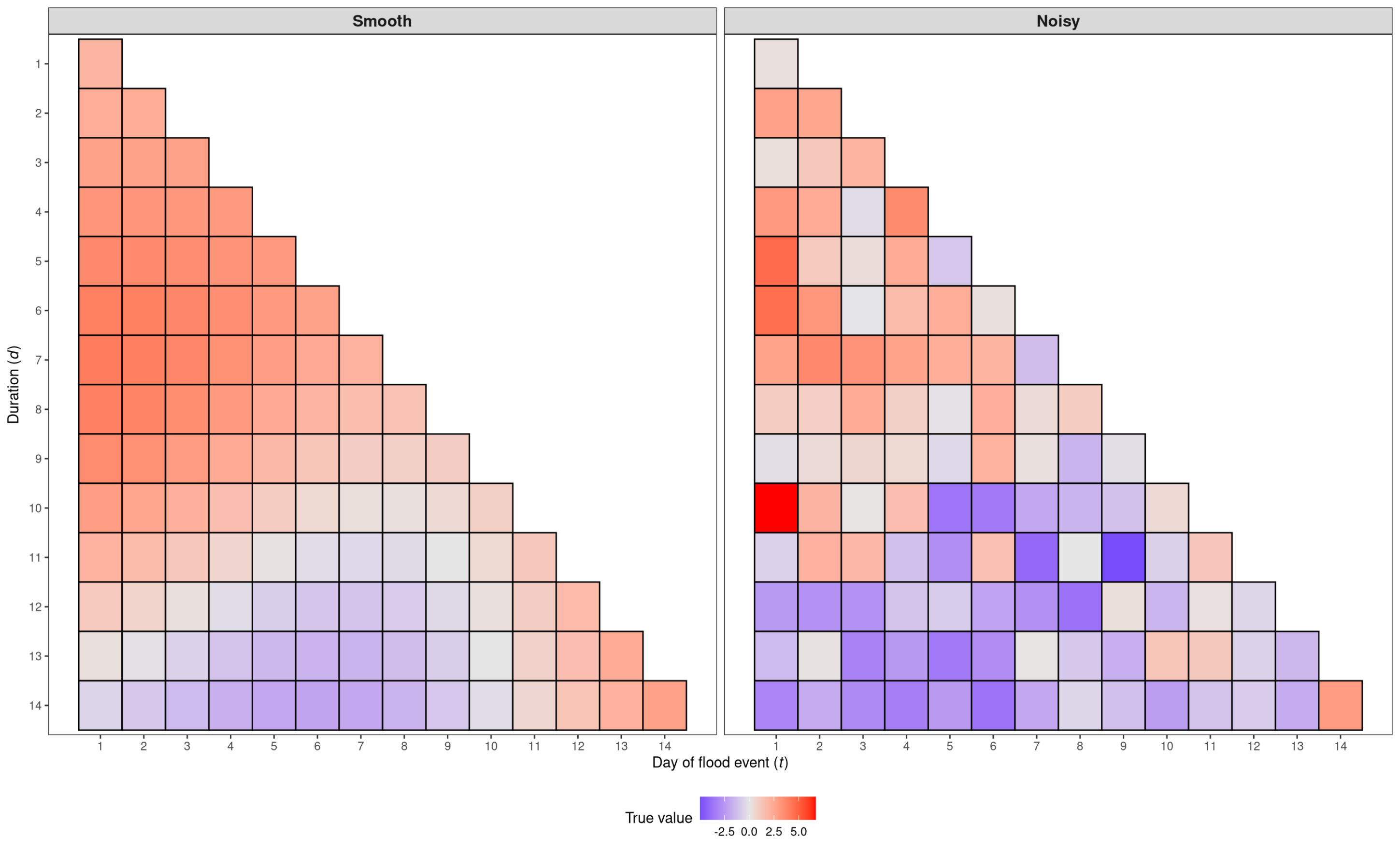}
    \caption{Ground truth $\beta_{d,t}$ duration-day coefficients in the smooth and noisy scenarios.}
    \label{ground_truth}
\end{figure}

We conducted a simulation study to assess the performance of the EDVCM and compare to conventional approaches for handling exposures of varying duration. We used a random subset of the flood exposure data contained in our Medicare data application to obtain realistic data distributions. In these data, the maximum flood duration is $D=14$ days. 
We generated true $\boldsymbol{\beta}$ starting using a two-dimensional thin-plate regression spline with five basis functions and then added random noise at two levels to create smooth and noisy coefficient surfaces for the simulation scenarios. Specifically, we added Gaussian noise where the variance was equal to 25\% (smooth) or 100\% (noisy) of the underlying spline surface variance. Figure \ref{ground_truth} shows the true values of the generated coefficients. This data-generating mechanism produces coefficients that are not strictly smooth across duration or exposure day, allowing the simulations to better reflect real-world variability. It also avoids generating data directly from the GP prior used in the model, enabling a more robust evaluation of model performance under misspecification. 
Then, for each scenario, we generated $n_{sim} = 5000$ replicate datasets from the multinomial data generating process. Because our primary goal is to assess estimation of $\boldsymbol{\beta}$, we exclude time-varying covariates in the simulation setup. 

On each simulated dataset, we fit the EDVCM and two  ``conventional''  comparator methods. First, we apply a Bayesian comparator method in which we assign independent standard normal priors on each element of $\boldsymbol{\beta}$. Second, for each duration separately, we fit a frequentist generalized linear model (GLM) to obtain duration-day coefficient estimates. 
For each method, we compute the percent bias, the root mean square error (RMSE), and the 95\% credible interval (CI) coverage for each $\beta_{d,t}$.

In our main simulation study, floods of all durations in $[1,D]$ are observed in the data. In a secondary set of simulations, we evaluate the performance of our model when floods of certain durations are not present in the data. For these simulations, we use the same simulated data as in our main simulation study, but remove all floods of durations 4, 7, and 11 days from the observed data. In another set of secondary simulations, we include lag days in the data generating process and assess the EDVCM's performance in estimation of lagged (post-flood) effects. More details and results for secondary simulations are given in Appendix \ref{more_sim}. 



\begin{figure}[]
    \includegraphics[width=1\textwidth]{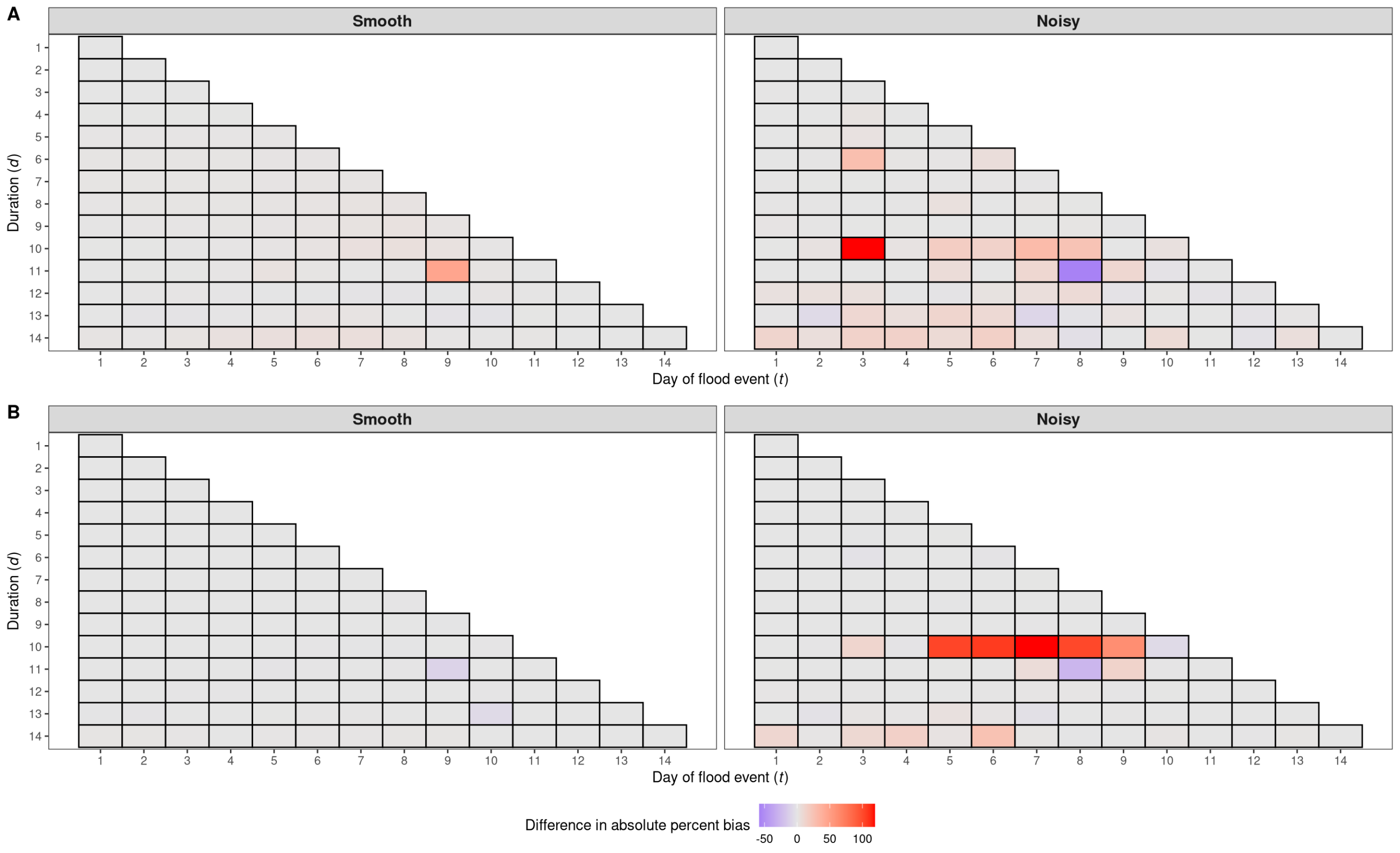}
    \caption{Panel A (first row) shows the difference in simulated absolute percent bias for coefficient estimation using the EDVCM vs independent standard normal priors on $\boldsymbol{\beta}$. Panel B (second row) presents the same quantity, but comparing the EDVCM to a frequentist GLM. Positive values, shown by red shading, indicate greater bias in the comparator method relative to the EDVCM.}
    \label{diff_in_bias}
\end{figure}

We observed that percent bias in the coefficient estimates from the EDVCM increased as the noise level increased, going from –0.54\% (9.35\%) in the smooth data to –2.06\% (20.24\%) in the noisy data. Figure~\ref{diff_in_bias} shows the difference in absolute percent bias of coefficient estimates, computed as the comparator method minus the EDVCM, for coefficient estimates from the Bayesian approach with independent normal priors (panel A) and from a frequentist GLM (panel B). The bias in the coefficient estimates from the EDVCM was generally smaller than with independent normal priors and with a frequentist GLM, especially in the noisy coefficient setting. 

\begin{figure}[]
    \includegraphics[width=1\textwidth]{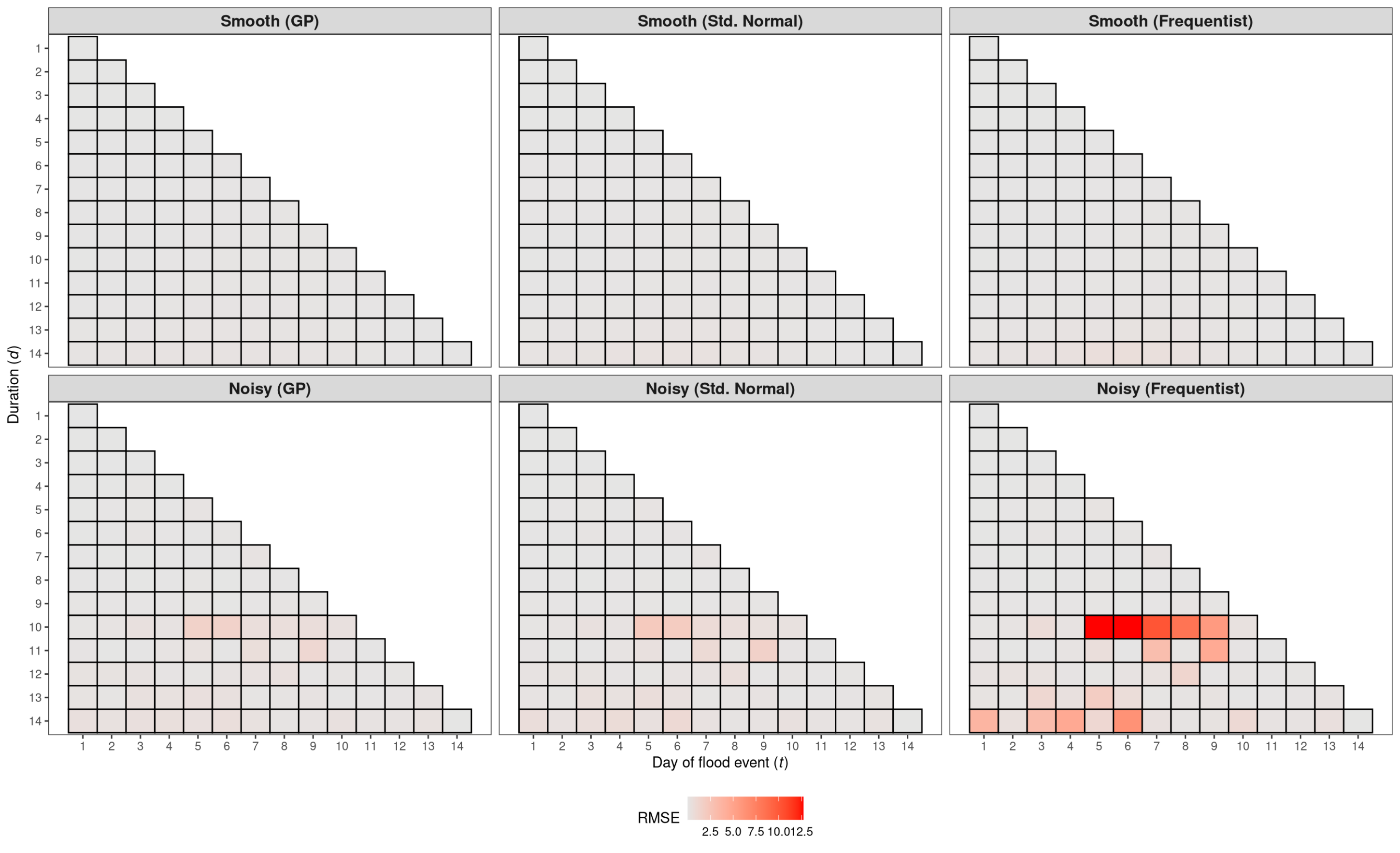}
    \caption{Simulated RMSE for three models: EDVCM (first column), the Bayesian comparator with independent standard normal priors (second column), and a frequentist GLM (third column).}
    \label{all_rmse}
\end{figure}

\begin{figure}[]
    \includegraphics[width=1\textwidth]{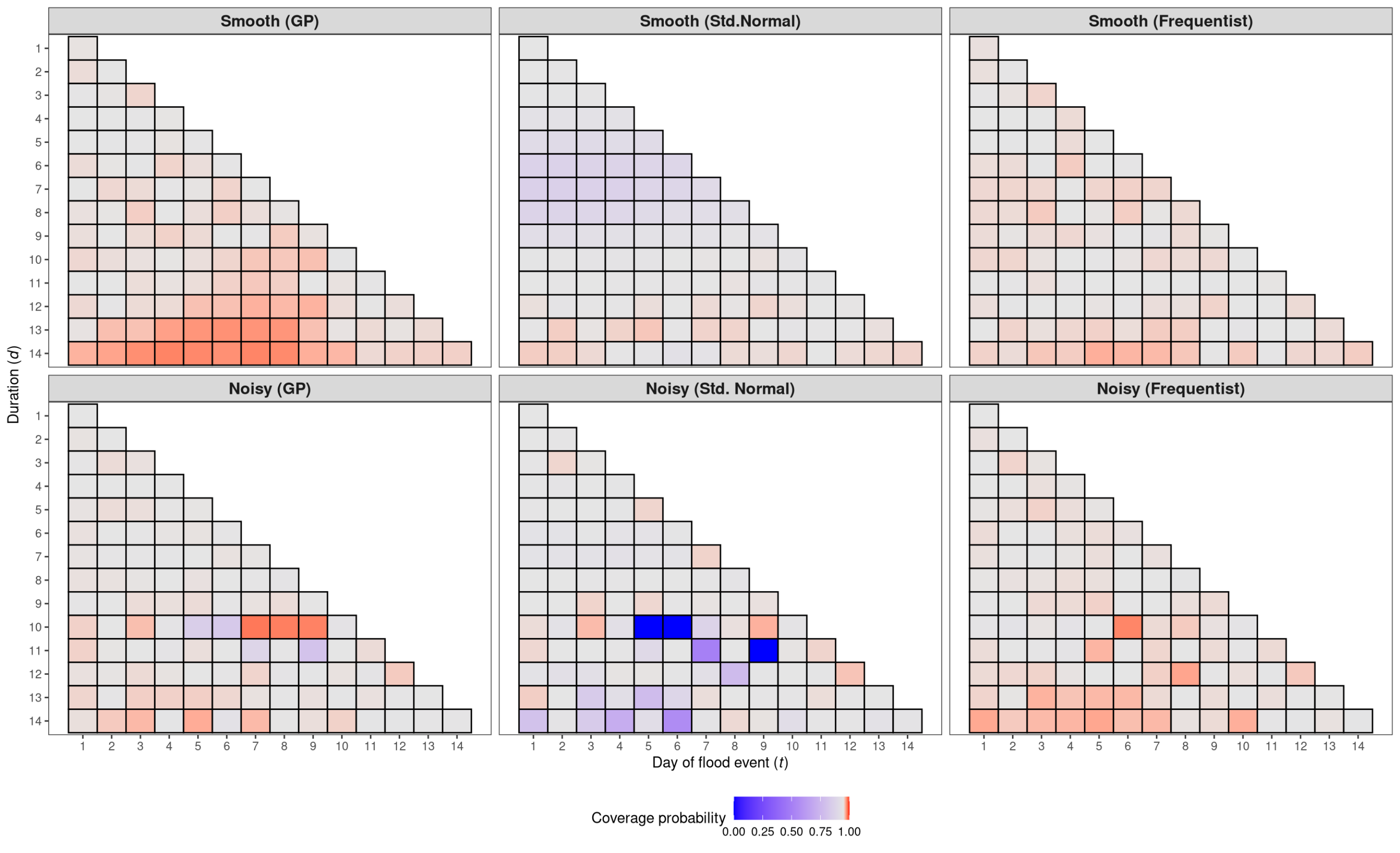}
    \caption{Simulated coverage probability for the EDVCM (first column), the Bayesian comparator with independent standard normal priors (second column), and a frequentist GLM (third column). Gray shading indicates coverage equal to 0.95 while red indicates coverage above 0.95 and blue indicates coverage below 0.95.}
    \label{coverage}
\end{figure}

Figures \ref{all_rmse} and \ref{coverage} show the RMSE and 95\% credible interval coverage, respectively, across simulated datasets for each coefficient estimate from the EDVCM (first column), the Bayesian comparator with independent standard normal priors (second column), and a frequentist GLM comparator (third column).
The EDVCM shows a lower RMSE for each coefficient than the Bayesian approach with independent standard normal priors and the frequentist approach. In general, we observed higher amounts of error in noisy versus smooth coefficient surfaces. 
The EDVCM and frequentist GLM generally achieved nominal or near-nominal coverage for most parameters. However, the model with independent standard normal priors significantly under-covered for many of the coefficients.

\section{Application}
The study covered 85 distinct flood events during 2000-2016 and exposed 2,872 unique counties in the contiguous US. We included flood-exposed counties that experienced between 1 and 10 days of flooding, excluding exposed counties with flood durations beyond this range due to both sparse data and the additional computational burden associated with modeling them. All durations between 1 and 10 days were observed in the data, with the median county-level average duration across flood events equal to 4.4 days (IQR: 3., 5.5) (Figure \ref{spatial_trends}, Panel B). Flood exposures were most common in the Mississippi and Ohio River Valleys and along the Gulf and Southeastern Atlantic coasts (Figure~\ref{spatial_trends}, Panel A). Across the study period, the flood events occurred most frequently between March (13 events) and September (12 events). Our analytic dataset included a total of 165,976 hospitalizations for musculoskeletal system diseases.  
We fitted the EDVCM to daily hospitalization counts. We adjusted for the time-varying covariates described in Section~\ref{sec:data} and did not include lag days in this analysis.  

\subsection{Results}
\begin{figure}[]
    \centering
    \includegraphics[width=1\textwidth]{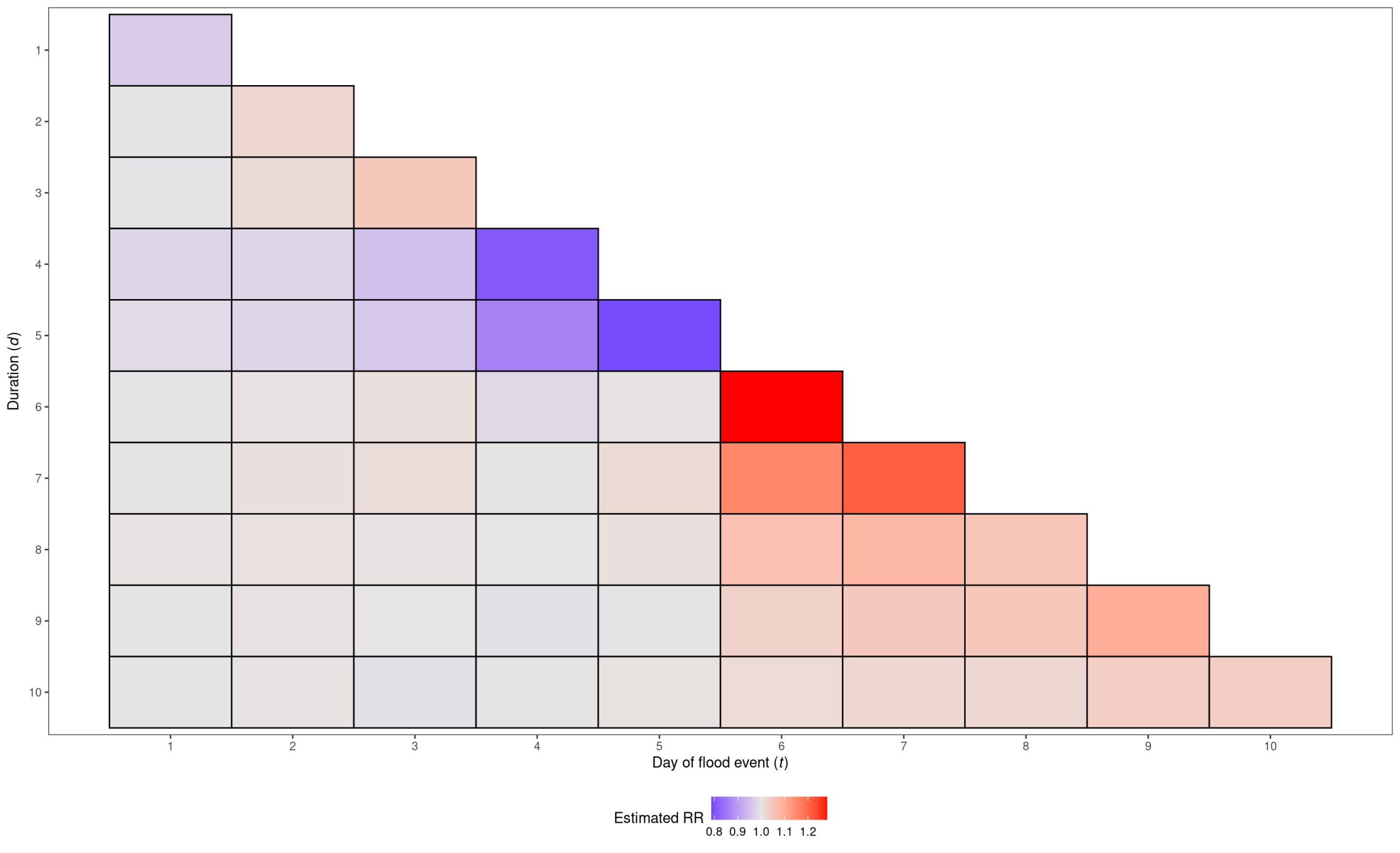}
    \caption{Duration- and day-specific flood-attributable hospitalization rate ratio estimates, $\text{exp}(\beta_{d,t})$, for musculoskeletal system hospitalizations.}
    \label{musc_betas}
\end{figure}

The results of our analysis for musculoskeletal disease hospitalizations are shown in Figures \ref{musc_betas} and
\ref{musc_dir}, and are the focus of our discussion here. 
The estimated associations between the time-varying covariates and hospitalization for each cause are given in Appendix \ref{assoc}.

In what follows, we present the rate ratio point estimates, $\text{exp}(\widehat{\beta}_{d,t})$, with the corresponding 95\% credible interval. In Figure \ref{musc_betas}, we observed decreased musculoskeletal system hospitalization rates during floods of duration one, four, and five days where the largest protective effect was on day 5 of a 5-day flood (0.79 [95\% CI: 0.77, 0.81]). In contrast, we observed harmful effects for floods of lengths 2-3 and 6+ days. Adverse effects of flood exposure on musculoskeletal-related hospitalizations were concentrated in the latter quarter of days for each of these durations, indicative of a potential window of vulnerability. The maximum rate ratio, which indicates a statistically significant increase in hospitalization rates, was observed on the last day of 6-day flood (1.28; 95\% CI: 1.25, 1.32). Floods of longer duration showed higher peak effect estimates for musculoskeletal system hospitalizations (1.21 [95\% CI: 1.17, 1.25] for a 7-day flood) beyond that of floods of shorter length (1.02 [95\% CI: 1.00, 1.05] for a 2-day flood). Additionally for each duration, the largest magnitude effect size was on the last day of the flood event. We identified that nine duration-day effect estimates had CIs excluding the null value, all of which occurred on the final day of the flood event (Figure \ref{musc_dir}). Three of these effects revealed decreased hospitalization rates (durations 1, 4, 5) while six revealed increased hospitalization rates (durations 2, 3, 6-9) for musculoskeletal system diseases. 

Table \ref{cumulative_results} gives the cumulative rate ratios for flood exposure on hospitalization for musculoskeletal system diseases and the corresponding 95\% CI for each flood duration. The CIs for the cumulative rate ratios for all durations crossed 1 (the null value); however, we found that floods of length seven days had the largest adverse impact on musculoskeletal system diseases while floods of length five days showed the largest protective effect.

\begin{table}[]
    \centering
    \begin{tabular}{|c|c|}
    \hline
        \textbf{Duration} & \textbf{Cumulative rate ratio (95\% CI)} \\
        \hline 
         1 & 0.96 (0.95, 0.98) \\
         2 & 1.02 (0.95, 1.11) \\
         3 & 1.04 (0.90, 1.19) \\
         4 & 0.92 (0.78, 1.08) \\
         5 & 0.91 (0.76, 1.07) \\
         6 & 1.07 (0.92, 1.26) \\
         7 & 1.08 (0.92, 1.28) \\
         8 & 1.04 (0.90, 1.22) \\
         9 & 1.04 (0.89, 1.23) \\
         10 & 1.03 (0.89, 1.20) \\
         \hline
    \end{tabular}
    \caption{Cumulative rate ratio estimate of flood exposure on musculoskeletal system hospitalizations for each duration with corresponding 95\% credible interval.}
    \label{cumulative_results}
\end{table}

\section{Discussion}
Flood events are expected to increase in severity and frequency due to anthropogenic climate change and older Americans are particularly vulnerable to the negative impacts of flood exposure during and following the flood event. While large-scale cohort studies have recently been conducted on flood exposure and adverse health outcomes, including cause-specific hospitalization and mortality \citep{yang2023mortality,lynch2025large,aggarwal2025severe}, current methods do not take into account the heterogeneity of effects that is likely to arise due to the different durations of flood events. Previous studies have typically modeled daily or monthly aggregations of flood exposure without explicitly accounting for flood duration, treating multi-day events as sequences of exposure days or as a single exposure period \citep{yang2023mortality, lynch2025large, aggarwal2025severe}. However, this approach results in a loss of information that could instead be leveraged to characterize heterogeneity in flood-related health impacts. This paper presents the EDVCM to estimate joint effects of daily exposure on rates of hospitalization, as well as their interactions with the length of the flood event while minimizing time-invariant confounding. In doing so, we are able to understand whether exposure day-specific flood effects vary with the length of the flood event and whether the length of the flood event results in different critical windows. Our work is motivated by the need for a more comprehensive and flexible analytic approach for accommodating environmental exposures of varying duration and characterizing duration-driven effect heterogeneity.

We utilized nationwide Medicare data from 2000 to 2016 and satellite-based flood maps to characterize heterogeneity in flood-related hospitalizations in older adults. For musculoskeletal system hospitalizations, we observed that the largest effect sizes were found in the later days of the flood event for each duration, indicative of critical windows of vulnerability. Floods that span 6-9 days had the largest positive associations with hospitalization rates, while floods that span fewer days revealed decreased hospitalization rates for this particular disease cause. This suggests that day-specific estimates do vary with the duration of the flood event as hypothesized. 

Because a substantial proportion of musculoskeletal system hospitalizations are for chronic sub-causes, individuals may delay care during shorter flood events and the earlier days of longer flood events due to the imminent risks associated with traveling to a hospital \citep{radcliff2018model}. This may lead to increased hospitalization rates in the later days of longer flood events as people cannot wait longer before they need help, consistent with what we observed. This pattern suggests that duration of flood event can serve as a proxy for severity. Previous work that collapsed all flood exposure days together into a single ``exposure period'' found a decreased rate of hospitalization for musculoskeletal system diseases during the exposure period \citep{aggarwal2025severe}. However, with the duration-day effect estimates generated here that provide a higher level of detail, we can see that this could have been driven by decreased rates of hospitalization in more common, shorter flood events, when individuals are able to postpone care for a short period of time. The collapsed exposure period masked increases in hospitalization that occur during the later days of longer (and likely more severe) floods.

While our work contributes a novel statistical model for understanding the health consequences of flood events, our study has limitations. Flooding is a highly localized event, but we aggregated flood exposures to the county-level for linkage with Medicare records. As a result, exposure misclassification is possible since not everyone living in an exposed county experiences exposure directly. However, flood events can impact the health of the broader community including those whose homes were not flooded, and our approach allows us to capture these effects, including changes that result from disruptions to health systems and healthcare access. Secondly, we utilized a robust self-matched study design that controls for time-invariant factors across counties and seasonality while adjusting for longer-term time trends in meteorology and pollution. However, we cannot rule out residual confounding as in any observational study. Third, our model assumes common effects across space. An area of future work would be to allow coefficients to vary across space as well in order to provide even more highly resolved insights. An additional limitation of our model is that its dimensionality grows quadratically with the maximum observed flood duration, which may be computationally burdensome in settings with extremely long or outlier events. Lastly, Poisson models assume that the mean and variance are equal; however, overdispersed counts in Medicare hospitalization data are plausible, thus distributions that account for this phenomenon may lead to improved model performance in real data.  

\section*{Declarations}

\noindent\textbf{Funding.} This work was supported by the Harvard Data Science Initiative and the Harvard Milton Fund, as well as National Institutes of Health grants K01ES032458, P30ES000002, T32CA009337, and T32ES007142. 

\noindent\textbf{Acknowledgements.} The computations in this paper were run on the FASRC Cannon cluster and FAS Secure Environment as well as the Regulated Data Environment, supported by the Faculty of Arts and Sciences Division of Science Research Computing Group and the Office of the Vice Provost for Research at Harvard University.

\noindent\textbf{Conflicts of interest.} The authors have no conflicts of interest to disclose. 

\noindent\textbf{Ethics approval.} This study was approved by the Harvard Longwood Campus Institutional Review Board, IRB20-1910 Tropical Cyclones. 

\noindent\textbf{Availability of data and materials.} Flood data is available via Harvard Dataverse \citep{flood_data}. Medicare enrollee data are publicly accessible, upon purchase after an application process, from the CMS \citep{cms}. 

\noindent\textbf{Code availability. } Code for analysis and visualizations presented in this manuscript is available at \nolinkurl{https://github.com/sarika1999/EDVCM}.

\clearpage
\printbibliography
\clearpage

\counterwithin{figure}{section}
\counterwithin{table}{section}

\renewcommand{\thesection}{S}

\doublespacing
\section{Supplementary materials}

\subsection{Priors on hyperparameters and remaining model parameters}\label{hyperpriors}
We considered several weakly informative priors for each hyperparameter in the simulation study, as well as for the model parameters corresponding to the vector of time-varying covariates used in the data application. We parameterize the marginal variance as $\sigma^2_{\beta}$, where $\sigma_{\beta}$ denotes the marginal standard deviation and is assigned a prior. We adopt the same parameterization for $\sigma_{\theta}$, placing the prior on the corresponding standard deviation. Table \ref{hyperprior_summary} lists the final set of priors used for each parameter. 
\begin{table}[h!]
\begin{center}
\begin{tabular}{|c|c|c|}
\hline
\textbf{Parameter} & \textbf{Simulation} & \textbf{Application} \\ \hline
$\sigma_{\beta}$ & lognormal(0.5*log(0.3),0.1) & t(3,0,1) \\ \hline
$\phi$ & lognormal(log(0.3),0.2) & lognormal(0, 0.6) \\ \hline
$\tau$ & lognormal(log(0.3),0.2) & lognormal(0, 0.6) \\ \hline 
$\sigma_{\theta}$ & lognormal(0.5*log(0.3),0.1) & N/A \\ \hline
$\gamma$ & lognormal(log(0.3),0.2) & N/A \\ \hline
$\eta$ & lognormal(log(0.3),0.2) & N/A \\ \hline 
$\zeta$ & N/A & N(0,100) \\ \hline
\end{tabular}
\caption{Priors on parameters}
\label{hyperprior_summary}
\end{center}
\end{table}
\vspace{-30pt}
\subsection{Cumulative effects derivation}\label{cumulative_deriv}
Let $\lambda_i(1)$ be the total number of events on days $t = 1, \dots, d$ under flood exposure ($A_i = 1$) and $\lambda_i(0)$ be the total number of events on days $t = 1, \dots, d$ absent flood exposure ($A_i = 0$). We are interested in the cumulative rate ratio for a particular duration: $$\Delta_{d} = \frac{\mathbb{E}[\lambda_i(1)|d(i)= d]}{\mathbb{E}[\lambda_i(0)|d(i)=d]}$$
\noindent\textbf{Case 1: No time-varying covariates}\\
Under the model specification given in Equation (\ref{poisson_model}) with $L_i = 0$ and $\zeta'z_{i}$, 
$$\begin{aligned}
\Delta_{d} = \frac{\mathbb{E}[\lambda_i(1)|d(i)=d]}{\mathbb{E}[\lambda_i(0)|d(i)=d]} &= \frac{\sum_{i:d(i)=d, t(i) \leq d} P_i \text{exp}(\alpha_0 + \alpha_{s(i)} + \beta_{d(i),t(i)})}{\sum_{i:d(i)=d, t(i) \leq d}P_i \text{exp}(\alpha_0 + \alpha_{s(i)})} \\
&= \frac{P_i\text{exp}(\alpha_0)\text{exp}(\alpha_{s(i)})\sum_{i:d(i)=d, t(i) \leq d} \text{exp}(\beta_{d(i),t(i)})}{d\times P_i\text{exp}(\alpha_0)\text{exp}(\alpha_{s(i)})}\\
&= \frac{1}{d}\sum_{i:d(i)=d, t(i) \leq d} \text{exp}(\beta_{d(i),t(i)})
\end{aligned}$$

\vspace{5pt}

\noindent \textbf{Case 2: Time-varying covariates}\\
Under the model specification given in Equation (\ref{poisson_model}) with $L_i = 0$,
$$\begin{aligned}
\Delta_{d} = \frac{\mathbb{E}[\lambda_i(1)|d(i)=d]}{\mathbb{E}[\lambda_i(0)|d(i)=d]} &= \frac{\sum_{i:d(i)=d, t(i) \leq d} P_i\text{exp}(\alpha_0 + \alpha_{s(i)} + \beta_{d(i),t(i)} + \zeta'z_{i})}{\sum_{i:d(i)=d, t(i) \leq d} P_i\text{exp}(\alpha_0 + \alpha_{s(i)} + \zeta'z_{i})}\\
&= \frac{P_i\text{exp}(\alpha_0)\text{exp}(\alpha_{s(i)})\sum_{i:d(i)=d, t(i) \leq d}\text{exp}(\beta_{d(i),t(i)})\text{exp}(\zeta'z_{i})}{P_i\text{exp}(\alpha_0)\text{exp}(\alpha_{s(i)})\sum_{i:d(i)=d, t(i) \leq d}\text{exp}(\zeta'z_{i})} \\
&= \frac{\sum_{i:d(i)=d, t(i) \leq d}\text{exp}(\beta_{d(i),t(i)})\text{exp}(\zeta'z_{i})}{\sum_{i:d(i)=d, t(i) \leq d}\text{exp}(\zeta'z_{i})}
\end{aligned}$$

\subsection{Additional simulation study results}\label{more_sim}

\begin{figure}[h!]
    \includegraphics[width=1\textwidth]{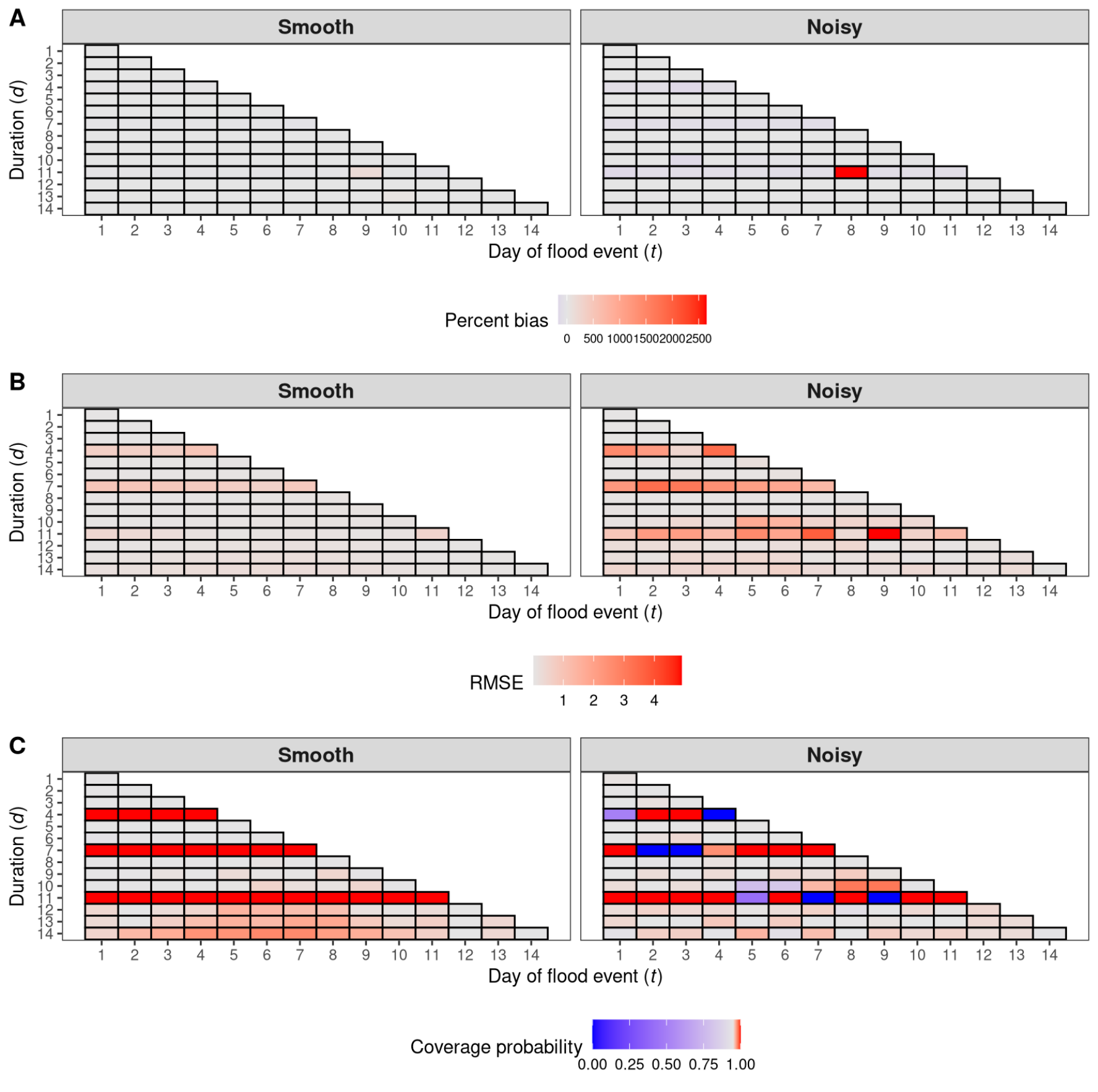}
    \caption{Simulation performance metrics (percent bias, root mean square error, and 95\% credible interval coverage) for coefficient estimates from the EDVCM when durations (d = 4, 7, 11) are not observed in the data.}
    \label{missingdur}
\end{figure}

In our first set of secondary simulations, the EDVCM yields larger biases (panel A) and RMSE (panel B) as well as worsened coverage probability (panel C) for coefficients corresponding to durations not present in the data (Figure \ref{missingdur}). This is especially evident in the noisy coefficient setting as compared to the smooth setting. Despite this, the EDVCM performs well across all metrics for coefficients corresponding to those durations that are observed in the data for both the smooth and noisy $\boldsymbol{\beta}$ scenarios. This phenomenon aligns with our expectations.

\begin{figure}[h!]
    \includegraphics[width=1\textwidth]{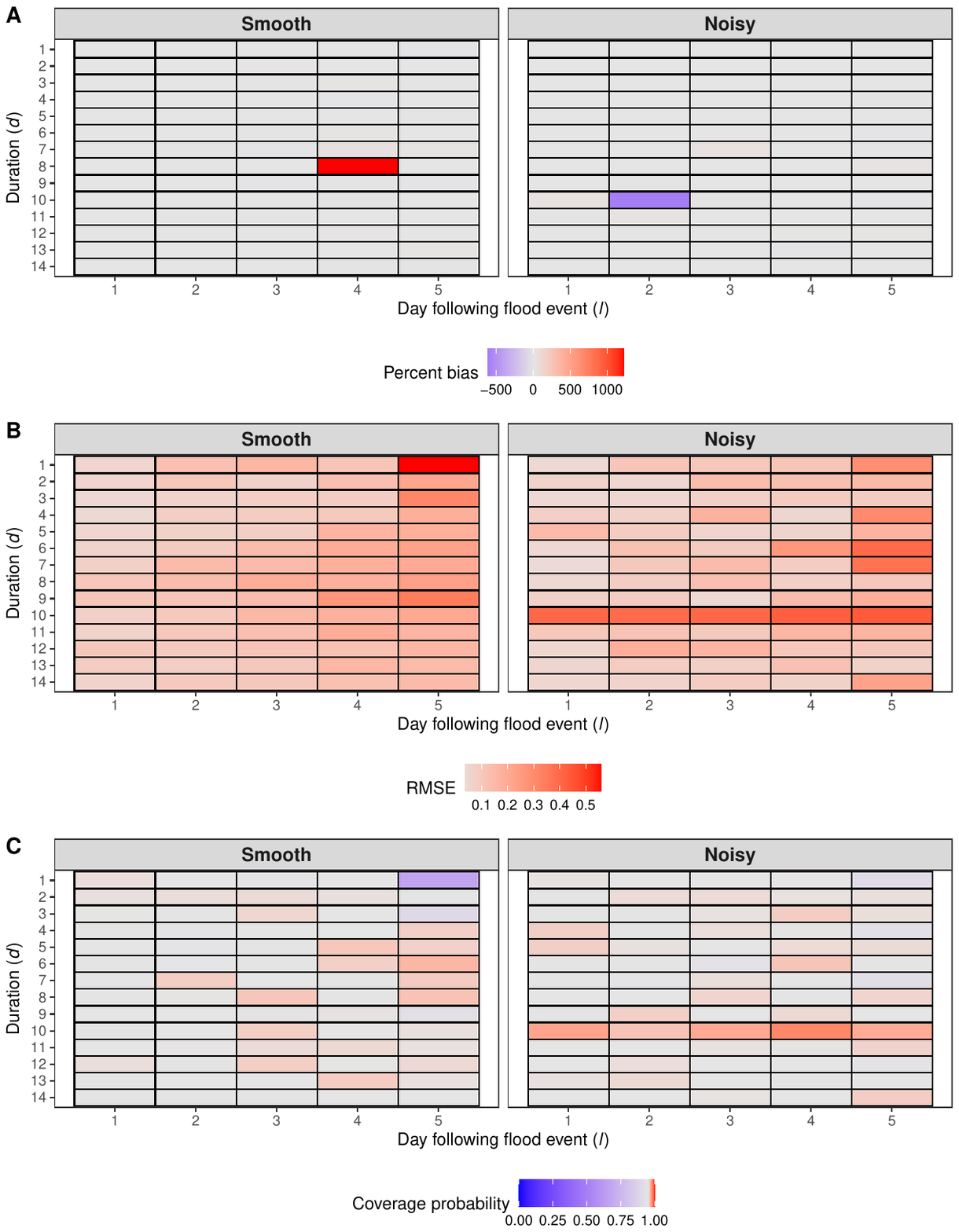}
    \caption{Simulation performance metrics (percent bias, root mean square error, and 95\% credible interval coverage) for lagged coefficient estimates (5 post-flood days) from the EDVCM.}
    \label{lagged_plot}
\end{figure}

In simulations that included post-flood effects, we considered 5 lagged day effects for each duration. We utilized a smooth (for smooth exposure effects) and noisy (for noisy exposure effects) lagged effects construction based on the same data-generating process described in \ref{simulation_sec}. We assigned a two-dimensional Gaussian process prior to the lagged effects (\ref{theta_kernel}) with hyperparameter priors given in \ref{hyperprior_summary}. Results are shown in Figure \ref{lagged_plot}. The percent bias (panel A) of the smooth and noisy lagged effects are minimal with the exception of some outliers. Due to a few extreme values, we also present the median percent bias for the smooth construction as -0.90\% (IQR: -2.64\%, 0.34\%) and for the noisy construction as -0.45\% (IQR: -1.48\%, 0.04\%). RMSE values (panel B) for both scenarios are similar, with earlier post-flood effects (lag days 1-2) having smaller magnitudes of error than later post-flood effects (lag days 3-4). Lastly, 95\% credible interval coverage (panel C) is achieved for most lagged coefficient estimates. 

\subsection{Additional application results}\label{more_app}
\begin{figure}[h!]
    \centering
    \includegraphics[width=1\textwidth]{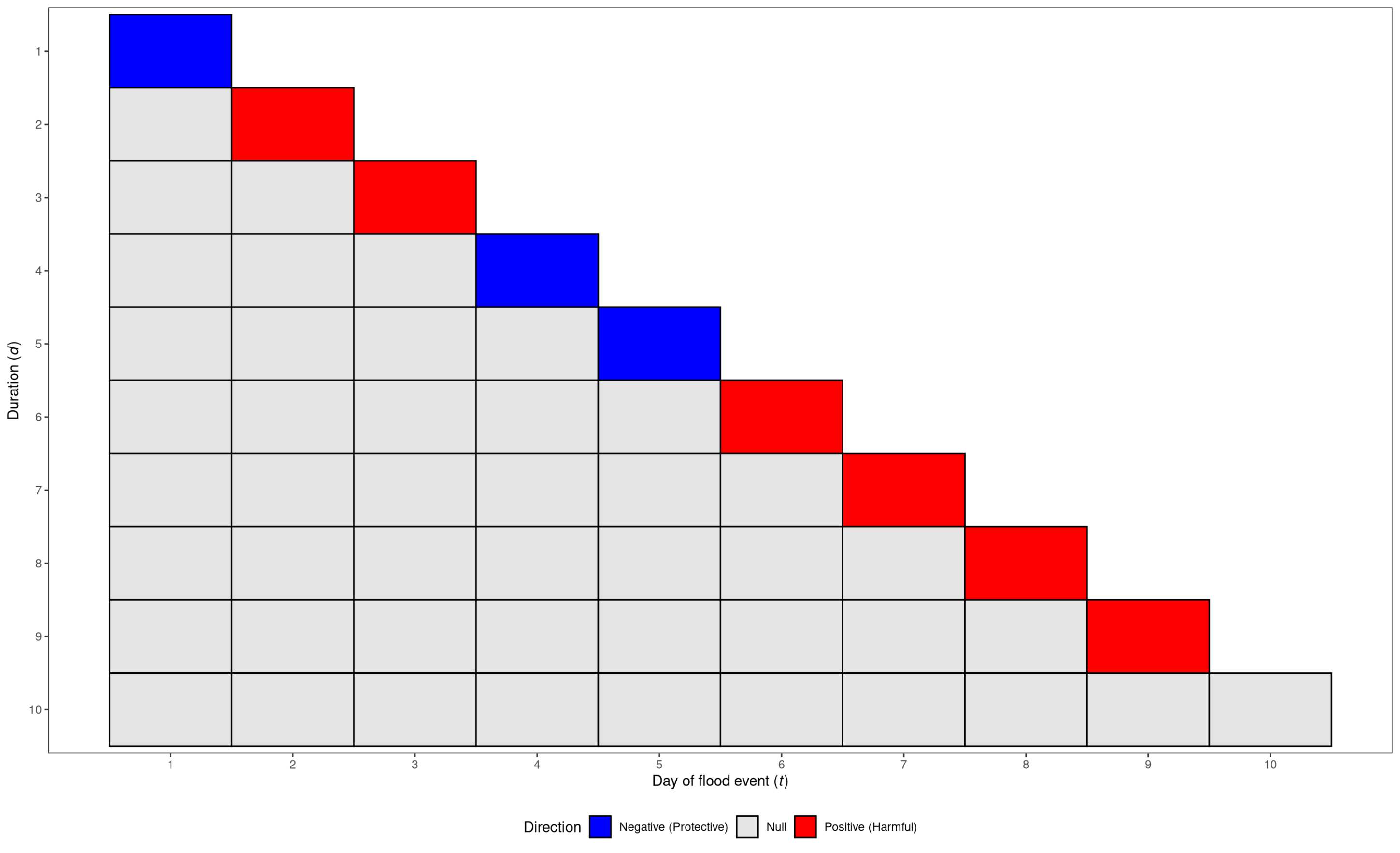}
    \caption{Direction of each $\beta_{d,t}$ coefficient for musculoskeletal system diseases based on the corresponding 95\% credible interval. With regards to flood exposure on hospitalizations, blue squares indicate protective effects, red squares indicate harmful effects, and gray indicates no effect.}
    \label{musc_dir}
\end{figure}

Figure \ref{musc_dir} shows the direction of each $\beta_{d,t}$ coefficient estimate, using the 95\% credible interval, for the impact of flood exposure on musculoskeletal system disease hospitalizations. 

\subsubsection{Time-varying covariates in the EDVCM}\label{assoc}
We adjust for time-varying meterological factors in our data application by including a spline with 3 degrees of freedom for each of the following covariates: air temperature, relative humidity, windspeed, fine particulate matter (PM2.5), ozone (O3), and nitrogen dioxide (NO2). Table \ref{confounder_assoc} gives their associations:

\begin{table}[h!]
    \centering
    \begin{tabular}{|c|c|c|c|}
    \hline
    Covariate & Estimate (95\% CI) \\
        \hline 
         Temperature(1) & -0.01 (-0.02, -0.01)  \\
         Temperature(2) & 0.09 (0.06, 0.13) \\
         Temperature(3) & -0.16 (-0.24, -0.09)  \\
         \hline
         Humidity(1) & 0.00 (0.00, 0.00)  \\
         Humidity(2) & -0.01 (-0.04, 0.02)  \\
         Humidity(3) & 0.07 (-0.24, 0.40)  \\
         \hline
         Windspeed(1) & 0.35 (0.27, 0.44)  \\
         Windspeed(2) & -0.96 (-1.37, -0.58)  \\
         Windspeed(3) & 0.63 (0.11, 1.17)  \\
         \hline
         PM2.5(1) & 0.04 (0.00, 0.08) \\
         PM2.5(2) & -0.09 (-0.25, 0.07)  \\
         PM2.5(3) & 0.01 (-0.18, 0.20)  \\
         \hline
         O3(1) & -0.09 (-0.10, -0.09) \\
         O3(2) & 0.39 (0.35, 0.43)  \\
         O3(3) & -0.48 (-0.56, -0.42) \\
         \hline
         NO2(1) & 0.40 (0.35, 0.45)  \\
         NO2(2) & -0.95 (-1.14, -0.76) \\
         NO2(3) & 0.52 (0.32, 0.72)  \\
         \hline
    \end{tabular}
    \caption{Estimated associations between time-varying covariates and musculoskeletal system disease hospitalizations, presented as point estimates with corresponding 95\% credible intervals.}
    \label{confounder_assoc}
\end{table}

\end{document}